\documentclass[]{article}
\usepackage{geometry}
\usepackage{epsf,amsmath,amssymb,graphicx,setspace,epstopdf}

\title{The von Neumann Entropy for Mixed States}
\author{$^{[1]}$ Jorge A. Anaya-Contreras, $^{[2,+]}$ H\'ector M.  Moya-Cessa,$^{[1,*]}$ Arturo Z\'u\~niga-Segundo\\ \\		$[1]$ Instituto Polit\'ecnico Nacional, ESFM, Departamento de F\'isica. Edificio 9,\\ Unidad Profesional Adolfo L\'opez Mateos, CP 07738 CDMX, Mexico\\
$[2]$ Instituto Nacional de Astrof\'{\i}sica, \'Optica y Electr\'onica, Calle Luis Enrique Erro 1,\\ Santa Mar\'{\i}a Tonantzintla, Puebla, 72840 Mexico}
\date{[+] hmmc@inaoep.mx, [*] azuniga@esfm.ipn.mx}
		
\begin{document}

\maketitle
%
%
\thispagestyle{empty}

\begin{abstract}
	The Araki-Lieb inequality is commonly used to calculate the entropy of subsystems when they are initially in pure states as this forces the entropy of the two subsystems to be equal after the complete system evolves. Then, it is easy to calculate the entropy of a large subsystem by finding the entropy of the small one.  To the best of our knowledge, there does not exist a way of calculating the entropy when one of the subsystems is initially in a mixed state. We show here that it is possible to use the Araki-Lieb inequality in this case and  find the von Neumann entropy for the large (infinite) system. We show this in the two-level atom-field interaction.
\end{abstract}

\section*{Introduction}
It is well-known that the atomic inversion for a two-level atom interacting with a quantized field suffers collapses and revivals of Rabi oscillations. The revivals may be considered as an indicator of the nature of the photon distribution of the initial field being inside the cavity. For instance, in the case that a squeezed field is considered, the atomic inversion shows so-called ringing revivals which tell give us information that such non-classical field was used as an initial state \cite{Satya,Vidiella}. On the other hand, decoherence plays an strong role in the purity of the states of quantum systems \cite{Phoenix3} and effects of the effects of time dependent coupling parameters on qubit-field interactions under decay processes have already been studied \cite{Abdel}.

One of the most important quantities to measure the entanglement between two subsystems is the von Neumann entropy \cite{Neumann}, which, together with the atomic inversion may give information about the generation of nonclassical states. For instance, if the entropy of a quantized field initially in a squeezed state is close to zero and the atomic inversion is in the collapse region, it is known that a superposition of squeezed states is generated \cite{Vidiella}.

On the other hand, if an initial coherent field is considered, it  produces approximately a superposition of coherent states at half the revival time \cite{Banacloche}.

However, if instead of an initial coherent state, a so-called Schr\"odinger cat is considered, the revival
of oscillations is divided by two, and the Rabi oscillations occur sooner \cite{Vidiella2}. Now, if a mixture of coherent states is initially considered, the revival of Rabi oscillations appear as in the case of a single coherent state.

The atomic inversion, then, can give us some information about the initial state of the field that, and, together with entropy can tell us if the initial field was in a pure coherent state or in a mixture of coherent states, because both produce the same atomic inversion \cite{Phoenix2,Barnett,Banacloche}.

The Araki-Lieb inequality \cite{Araki}
\begin{equation}
\mid S_A-S_B\mid\le S_{AB}\le S_A+S_B
\end{equation}
where $S_{AB}$ is the von Neumann entropy   of the total system  and $S_A$ and $S_B$ are the entropies of the subsystems A and B, respectively, is of great help when one needs to calculate the entropy of a subsystem if the two susbsystems are in pure states, because the total wavefunction is also in a pure state and unitary evolution keeps the evolved wavefunction pure. In other words, the fact that $S_{AB}=0$ produces that both entropies are equal, $S_A=S_B$, and then, by computing the entropy for the small system, let us say a two-level atom, allows us to know the entropy for the larger system, which we will consider by simplicity a quantized field.

Therefore, if in a non-dissipative interaction, one can generate initial pure states for both subsystems, A and B, after evolution their entropies will be equal. 

If we consider the subsystems to be a two-level atom and a quantized field we may be able to find the entropy for the (big system) field given by the entropy of the (small system) atom, provided they were initially defined by a wavefunction. But, Is it possible to calculate the entropy of the field when it is initially in a mixed state, namely a mixture of coherent states? We have tried to give an answer to this question but were only able to deliver a positive answer only for certain periods of time, but not for the complete evolution \cite{23}. This is because in such case the above triangle inequality seems to be useless and there is not a general answer.

Phoenix and Knight showed how the field entropy for an initial coherent state can be calculated analytically for initial pure states for the atom and the field without making use of the Araki-Lieb inequality. They were able to find eigenstates and eigenvalues of the field density matrix.   Phoenix applied later the same method to compute the entropy
for a field subject to decay. Calculations of this type are already complicated for initially pure states for both subsystems. 

In this contribution we believe that we have a final answer to the problem of finding the field entropy even though any of the subsystems is in a mixed state. In this case $S_{AB}\ne 0$ and therefore we can not say much about the subsystems entropies. However, we will introduce the idea of a virtual four-level atom and still will be able to use the Araki-Lieb inequality. The solution we provide may be easily generalized to atoms with more levels  or more complicated mixtures of atomic or field states. We will start by introducing our method which consists in the fact that, once we trace over the atomic basis in order to obtain a field density matrix, we use the concept of virtual (many-level) atoms that will be the key for our calculation. We finally analyze, as an example, the atom field interaction in some detail when different initial mixed states are considered, {\it i.e., } specifically when the Araki-Lieb inequality can not be used to obtain the entropy of the large system (field) in term of the small one (atom).

\section*{Schmidt decomposition}
The Schmidt decomposition \cite{10} is a useful mathematical tool that plays an important role 
in one of the key features of quantum mechanics, namely, the description of entanglement.

Let us consider a wave function $\mid\psi\rangle$ of an entangled state that describes the interaction between an $n$-level system (an atom, for simplicity) and an infinite-level system (a quantized field)
\begin{equation}
\label{SM1}
\mid\psi\rangle =\sum_{k=1}^{n}\mid\psi_{k}\rangle\mid a_{k}\rangle,
\end{equation}
where $\{ \mid\psi_{k}\rangle\}_{_{F}}$ and $\{ \mid a_{k}\rangle\}_{_{A}}$ are a set of (unnormalized) field  and atomic states, respectively, which satisfy, in general, the following conditions
\begin{equation}
\label{SM1.1}
\langle\psi_{i}\mid\psi_{j}\rangle\neq0 \,\, , \langle a_{i}\mid a_{j}\rangle=\delta_{ij} \,\, , \forall \,\,\, i,j \in{ \{1,2,....,n\}} \,.
\end{equation}
Schmidt decomposition \cite{10} states that there exist a couple of orthonormal bases $\{ \mid\Psi_{k}\rangle\}_{_{F}}$ and $\{ \mid A_{k}\rangle\}_{_{A}}$  and real, non-negative numbers, $\lambda_{k}$, such that 
\begin{equation}
\label{SM2}
\mid\psi\rangle=\sum_{k=1}^{n}\sqrt{\lambda_{k}}\mid\Psi_{k}\rangle\mid A_{k}\rangle.
\end{equation} 
Moreover, the following quantity 
\begin{equation}
\label{SM3}
\sum_{n=0}^{k}\mid\psi_{k}\rangle\langle\psi_{k}\mid=\sum_{n=0}^{k}\lambda_{k}\mid\Psi_{k}\rangle\langle\Psi_{k}\mid
\end{equation}
is an invariant for such entangled state.
\section*{Entropy associated to an $n$-level system: Mixed states}
Consider $\hat{\rho}_{{M}}$ a density operator for a mixed state, defined by
\begin{equation}
\label{SM4}
\hat{\rho}_{_{M}}=\sum_{k=0}^{n} \mid\psi_{k}\rangle\langle\psi_{k}\mid\,,
\end{equation}
with 
\begin{equation}
\label{SM4.1}
\sum_{k=1}^{n}\langle\psi_{k}\mid\psi_{k}\rangle=1\,.
\end{equation}
Because of the invariant(\ref{SM3}), it may be rewritten as
\begin{equation}
\label{SM5}
\hat{\rho}_{_{M}}=\sum_{k=0}^{n} \lambda_{k} \mid\Psi_{k}\rangle\langle\Psi_{k}\mid \,,
\end{equation}
and, as the wavefunctions $\{ \mid\Psi_{k}\rangle \}_{_{F}}$ are orthonormal, the von Neumann entropy (defined as $S=-Tr\{\hat{\rho} \ln \hat{\rho}\}$) may be easily found
\begin{equation}
\label{SM6}
S{_{M}}=-\sum_{k=0}^{n}\lambda_{k}\ln \lambda_{k}\,.
\end{equation}
In what follows, we will show that the entropy to the mixed state (field), equation 
(\ref{SM6}), is equal to the entropy associated to a virtual atom and will verify that this fact is consistent with the Araki-Lieb inequality \cite{Araki}. In order to achieve this, we consider the density operator for the composed virtual-atom-field pure state, equation (\ref{SM2}), 
\begin{equation}
\label{SM7}
\hat{\rho}=\sum_{j=1}^{n}\sum_{k=1}^{n}\sqrt{\lambda_{j}\lambda_{k}}\mid\Psi_{j}\rangle\mid A_{j}\rangle\langle A_{k}\mid\langle\Psi_{k}\mid\,,
\end{equation}
that, by tracing over the field states, produces the atomic density operator
\begin{equation}
\label{SM8}
\hat{\rho}_{_{A}}=\sum_{k=0}^{n}\lambda_{k}\mid A_{k}\rangle\langle A_{k}\mid\,,
\end{equation}
from which we can easily obtain the (virtual) atomic entropy
\begin{equation}
\label{SM10}
S_{_{A}}=-\sum_{k=0}^{n}\lambda_{k}\ln \lambda_{k}\,.
\end{equation}
In addition, if we trace  the total density matrix, Equation (\ref{SM7}), over the atomic states we find that the field entropy may be written as 
\begin{equation}
\label{SM11}
S_{_{F}}=-\sum_{k=0}^{n}\lambda_{k}\ln \lambda_{k}\,,
\end{equation}
{\it i.e.}, both entropies are equal, $S_{_{F}}=S_{_{A}}$ and they are also equal to the entropy associated with the mixed state (\ref{SM6}). Therefore, in order to find the entropy for a mixed state (\ref{SM4}), one may construct an associated {\it  virtual} atom, calculate then its entropy, and, by virtue of the Araki-Lieb inequality, associate such {\it atomic} entropy to the field mixed state. It is possible to use the Araki-Lieb inequality because the density matrix (\ref{SM7}) is precisely a density matrix for a pure state, making the total entropy of the composed system equal to zero. Moreover, although, both entropies, for field and virtual atom, are not zero, they are equal, not violating the Araki-Lieb inequality.

It is also important to stress here that the maximum value of $S_{_{M}}$ is $\ln n$, because the virtual atom will be maximally entangled when all the probability amplitudes, $\sqrt{\lambda_{k}}$, for $k=1,2,...,n$ reach the same value, this is, $\lambda_{k}=1/n$.

\section*{Interaction between a two-level and a quantized field}
In order to apply our findings, we use as an example the well-known  Jaynes-Cummings model \cite{1}, whose Hamiltonian reads  
\begin{equation}
\label{SM13}
\hat{H}_{I}=\lambda\left(\hat{a}^{\dagger}\sigma_{-}+\hat{a}\sigma_{+}\right)\,,
\end{equation}
\begin{figure}
\centering
\includegraphics[width=10cm]{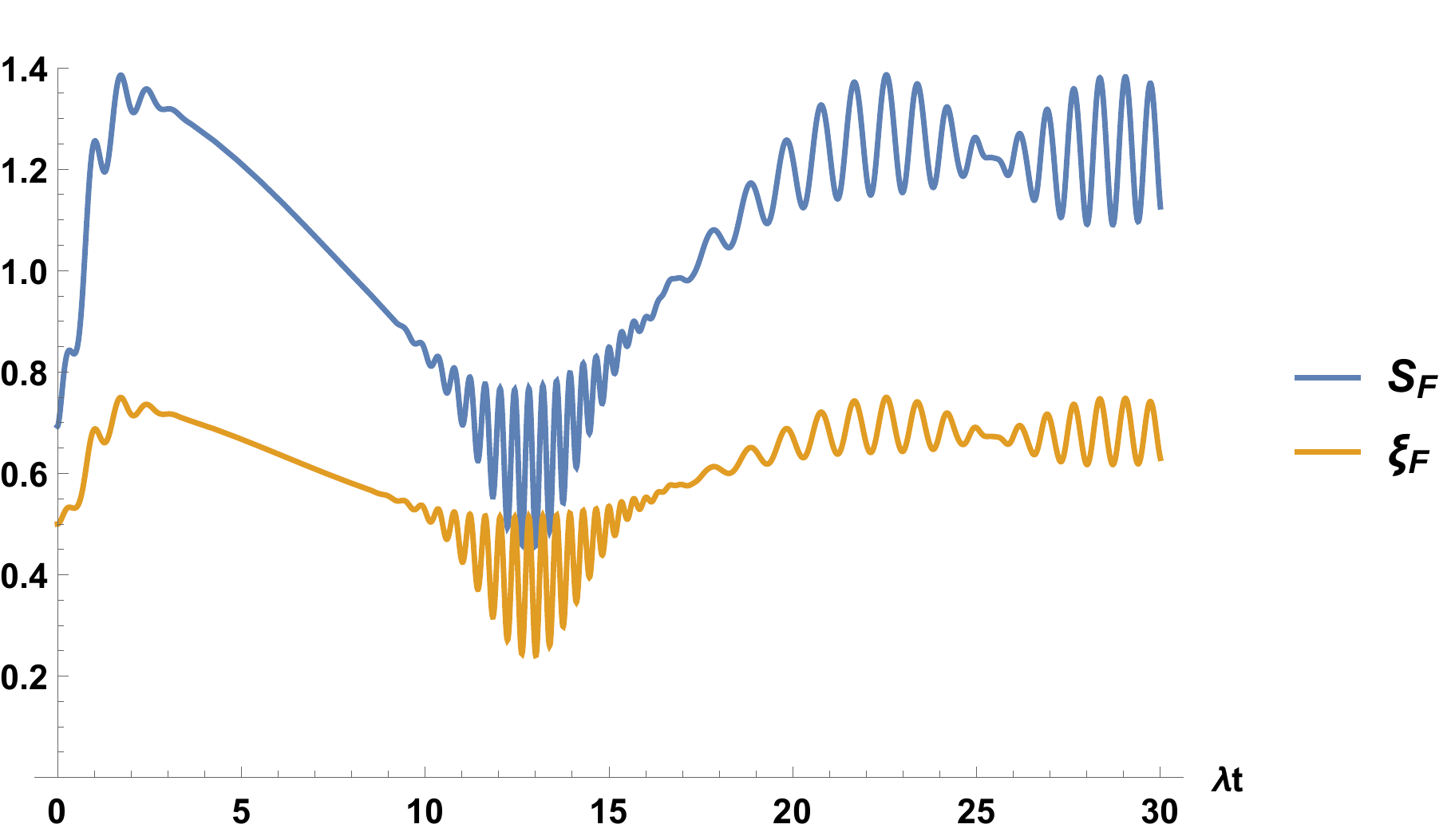} 
\caption{We plot the field entropy,  $S_{{F}}$, and the purity parameter, $\xi_{_{F}}$, as a function of $\lambda t$ for an initial field mixture of coherent states, with $\alpha=4$, $\beta=-4$, $\mathrm{C}=1/2$and the atom initially in its excited state.}
\label{SFPFSM}
\end{figure}
describes the interaction between a two-level atom and a quantized field in the rotating wave approximation. The interaction constant, $\lambda$, defines the rate at which the atom and the field exchange energy. The operator $\hat{a}$ and $\hat{a}^{\dagger}$ are the field annihilation and creation operators, respectively, while $\sigma_{-}$ and $\sigma_{+}$ are the atomic lowering and raising Pauli operators. It is not difficult to obtain the evolution operator for the Hamiltonian above, which reads \cite{Gerry}:
\begin{equation}
\label{SM14}
\hat{U}_I=
\begin{pmatrix}
\cos\left(\lambda t\sqrt{\hat{a}\hat{a}^{\dagger}}\right) & - i  \, \hat{V}\sin \left(\lambda t \sqrt{\hat{a}^{\dagger}\hat{a}} \right)\vspace{0.15cm} \\
- i \,  \hat{V}^{\dagger}\sin \left(\lambda t \sqrt{\hat{a}\hat{a}^{\dagger}} \right) &\cos\left(\lambda t\sqrt{\hat{a}^{\dagger}\hat{a}}\right)
\end{pmatrix}\,,
\end{equation}
with $\hat{V}$ and $\hat{V}^{\dagger}$ the London phase operators.

\subsection*{Initial field in a mixed state and  atom in an excited state}
First we consider the case of the field initially in a mixture of two coherent states and the atom in its excited state, {\it i.e.},  $\hat{\rho}(0)=\left(\mathrm{C}\mid\alpha\rangle\langle \alpha\mid+(1-\mathrm{C})\mid\beta\rangle\langle\beta\mid\right)\mid e\rangle\langle e\mid$, for which the evolved 
density matrix reads
\begin{equation}
\label{SM15}
\hat{\rho}=
\begin{pmatrix}
\mid\psi_{1}\rangle\langle\psi_{1}\mid+\mid\psi_{2}\rangle\langle\psi_{2}\mid & \mid\psi_{1}\rangle\langle\psi_{3}\mid+\mid\psi_{2}\rangle\langle\psi_{4}\mid \vspace{0.15cm} \\
\mid\psi_{3}\rangle\langle\psi_{1}\mid+\mid\psi_{4}\rangle\langle\psi_{2}\mid & \mid\psi_{3}\rangle\langle\psi_{3}\mid+\mid\psi_{4}\rangle\langle\psi_{4}\mid
\end{pmatrix}
\,,
\end{equation}

with
\begin{eqnarray}
\label{SM16}
\mid\psi_{1}\rangle&=&\sqrt{\mathrm{C}}\cos\left(\lambda t\sqrt{\hat{n}+1}\right)\mid\alpha\rangle\;,\nonumber\\ \mid\psi_{2}\rangle&=&\sqrt{1-\mathrm{C}}\cos\left(\lambda t\sqrt{\hat{n}+1}\right)\mid\beta\rangle\;, \nonumber\\
\mid\psi_{3}\rangle&=&-i\,\sqrt{\mathrm{C}}\hat{V}^{\dagger}\sin\left(\lambda t\sqrt{\hat{n}+1}\right)\mid\alpha\rangle\;,\nonumber \\ \mid\psi_{4}\rangle&=&-i\,\sqrt{1-\mathrm{C}}\hat{V}^{\dagger}\sin\left(\lambda t\sqrt{\hat{n}+1}\right)\mid\beta\rangle\;.
\end{eqnarray}
WE find the reduced atomic and field density operators by tracing over the field
\begin{equation}
\label{SM17}
\hat{\rho}_{_{A}}=
\begin{pmatrix}
\langle\psi_{1}\mid\psi_{1}\rangle+\langle\psi_{2}\mid\psi_{2}\rangle & \langle\psi_{1}\mid\psi_{3}\rangle^{*}+\langle\psi_{2}\mid\psi_{4}\rangle^{*}\vspace{0.2cm} \\ 
\langle\psi_{1}\mid\psi_{3}\rangle+\langle\psi_{2}\mid\psi_{4}\rangle & \langle\psi_{3}\mid\psi_{3}\rangle+\langle\psi_{4}\mid\psi_{4}\rangle 
\end{pmatrix},
\end{equation}
\begin{figure} 
\centering
\includegraphics[width=10cm]{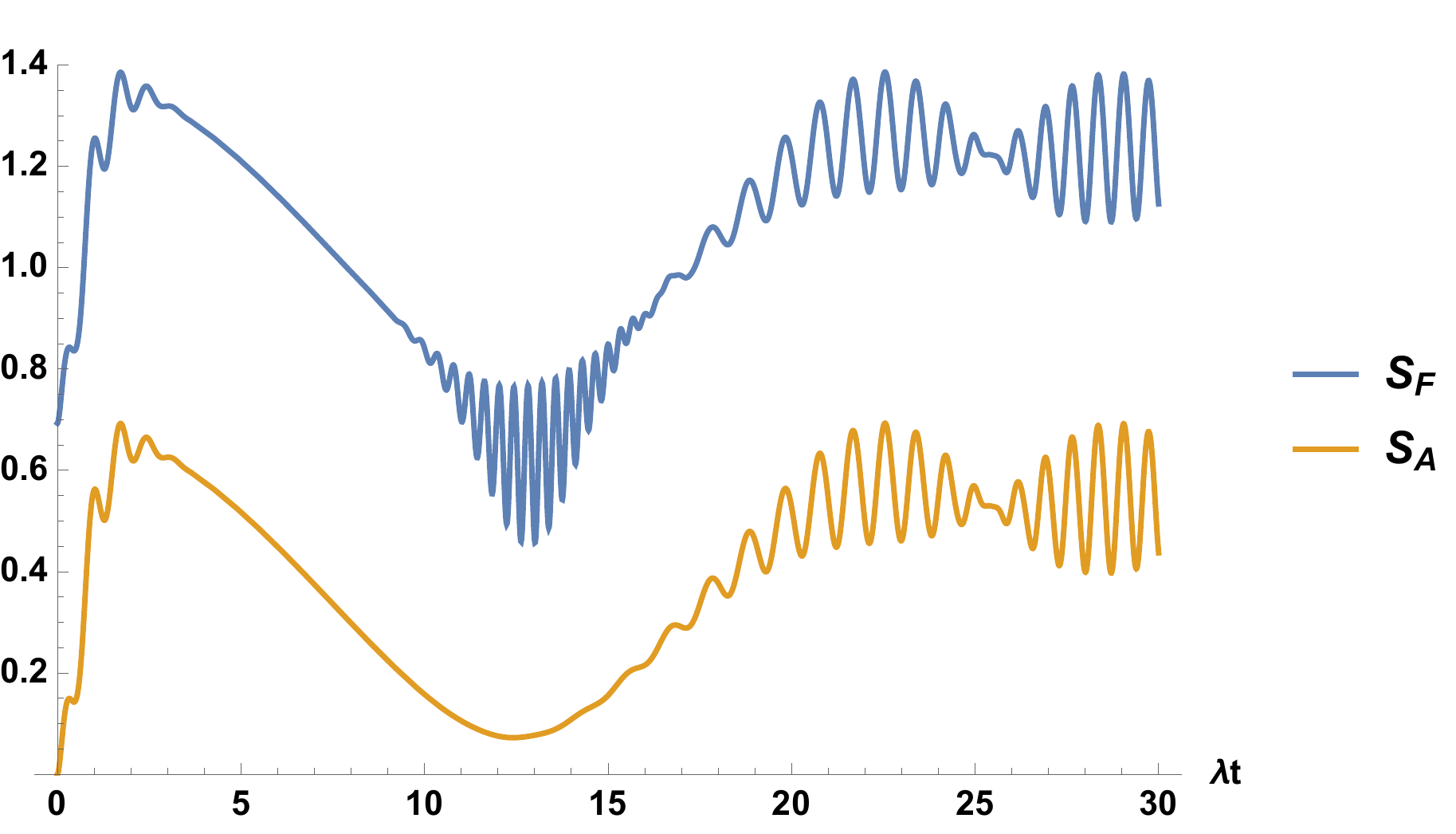} 
\caption{We plot the field entropy,  $S_{{F}}$, and the atomic entropy, $S_{_{A}}$, for the same parameters used in Figure 1.}
\label{SFSA}
\end{figure}
and the atomic basis
\begin{equation}
\label{SM18}
\hat{\rho}_{_{F}}=\mid\psi_{1}\rangle\langle\psi_{1}\mid+\mid\psi_{2}\rangle\langle\psi_{2}\mid+\mid\psi_{3}\rangle\langle\psi_{3}\mid+\mid\psi_{4}\rangle\langle\psi_{4}\mid,
\end{equation}
respectively. 

Note now that equation (\ref{SM18}) is precisely the invariant defined in equation (\ref{SM3}), but for a {\it virtual four-level atom}.

Because of equation (\ref{SM16}) the matrix elements of the density operator associated to the virtual four-level system are given by  $\mathrm{P}_{ij}=\langle\psi_{i}\mid\psi_{j}\rangle$.

Therefore, the entropy, $S_F$ and the purity parameter, $\xi_F=1-Tr\{\hat{\rho}_F^2\}$ may be easily calculated as
\begin{equation}
\label{SM21}
S_{_{F}}=-\lambda_{1}\ln \lambda_{1}-\lambda_{2}\ln \lambda_{2}-\lambda_{3}\ln \lambda_{3}-\lambda_{4}\ln \lambda_{4}
\end{equation}
and
\begin{eqnarray}
\label{SM22}
\xi_{_{F}}&=&1-\mid\mathrm{P}_{11}\mid^{2}-\mid\mathrm{P}_{22}\mid^{2}-\mid\mathrm{P}_{33}\mid^{2}-\mid\mathrm{P}_{44}\mid^{2}-2\mid\mathrm{P}_{12}\mid^{2} \nonumber\\
&{}&-2\mid\mathrm{P}_{13}\mid^{2}-2\mid\mathrm{P}_{14}\mid^{2}-2\mid\mathrm{P}_{23}\mid^{2}-2\mid\mathrm{P}_{24}\mid^{2}-2\mid\mathrm{P}_{34}\mid^{2}\;,
\end{eqnarray}
where the $\lambda_{j}$'s, are the solutions of the determinant equation
\begin{equation}
\label{SM23}
\det \begin{pmatrix}\mathrm{P}_{11}-\lambda &\mathrm{P}^{*}_{12}&\mathrm{P}^{*}_{13}&\mathrm{P}^{*}_{14}\\
\mathrm{P}_{12} &\mathrm{P}_{22}-\lambda&\mathrm{P}^{*}_{23}&\mathrm{P}^{*}_{24}\\
\mathrm{P}_{13} &\mathrm{P}_{23}&\mathrm{P}_{33}-\lambda&\mathrm{P}^{*}_{34}\\ 
\mathrm{P}_{14} &\mathrm{P}_{24}&\mathrm{P}_{34}&\mathrm{P}_{44}-\lambda\end{pmatrix}=0\,.
\end{equation}
Of course, the entropy for the {\it real} two-level atom is simply described by
\begin{equation}
\label{SM24}
S_{_{A}}=-\lambda_{+}\ln\lambda_{+}-\lambda_{-}\ln\lambda_{-}
\end{equation}
with
\begin{equation}
\label{SM25}
\lambda_{\pm}=\frac{1}{2}\pm\frac{1}{2}\sqrt{(\mathrm{P}_{11}+\mathrm{P}_{22}-\mathrm{P}_{33}-\mathrm{P}_{44})^{2}+4\mid\mathrm{P}_{13}+\mathrm{P}_{24}\mid^{2}}\,.
\end{equation}
\begin{figure} 
\centering
\includegraphics[width=13cm]{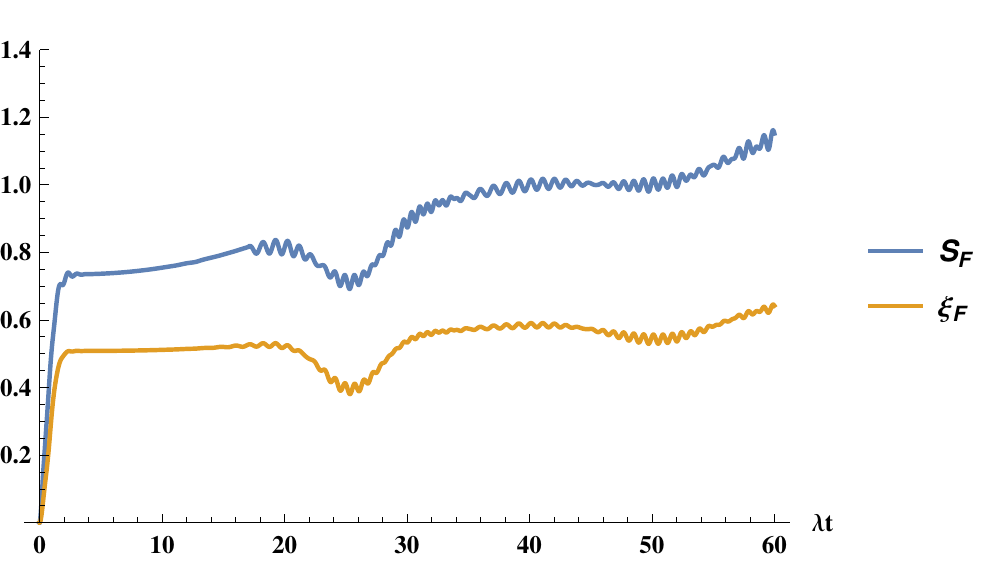} 
\caption{We plot the entropy, $S_{_{F}}$, and the field purity parameter, $\xi_{_{F}}$, for the atom initially in a mixture of ground and excited states, with $\mathrm{C}=1/2$ and the field initially in a coherent state with $\alpha=4$.}
\label{ASFPF05060}
\end{figure}
In figure \ref{SFPFSM} we plot the entropy, $S_{_{F}}$, and the purity parameter, $\xi_{_{F}}$, which as should be expected, show the same behaviour. We calculated in Reference \cite{23} the field entropy for certain periods of time, which agrees with the plots presented here, except for the period of time from about $\lambda t \approx 10$ to $\lambda t \approx 17$. Note that, although we are considering a two-level atom, the maximum of the entropy goes up to $\ln 4$, defined by our four-level virtual system.  

In figure \ref{SFSA} we plot the entropies for the field and the atom. They show different behaviour as it should be expected as they, because of the Araki-Lieb inequality are expected to be different. In fact, besides the difference by an amount $\ln 2$, the atomic entropy lacks the oscillation present in the field entropy for the period of time from about $\lambda t \approx 10$ to $\lambda t \approx 17$.   
\subsection*{Atom initially in a mixture of states and field in a coherent state} 
The formalism to calculate the entropy for mixed states can be extended to the case in which, not the field but the atom, is in a mixed state. In fact may be generalized to even more complicated cases, but we feel that it is enough to present this other case. Consider then an initial atom-field density matrix
$\hat{\rho}(0)=\left(\mathrm{C}\mid e\rangle\langle e\mid+(1-\mathrm{C})\mid g\rangle\langle g\mid\right)\mid\alpha\rangle\langle\alpha\mid$. In this case its evolution is described by
\begin{equation}
\label{SM26}
\hat{\rho}=
\begin{pmatrix}
\mid\psi_{1}\rangle\langle\psi_{1}\mid+\mid\psi_{2}\rangle\langle\psi_{2}\mid & \mid\psi_{1}\rangle\langle\psi_{3}\mid+\mid\psi_{2}\rangle\langle\psi_{4}\mid \vspace{0.15cm} \\
\mid\psi_{3}\rangle\langle\psi_{1}\mid+\mid\psi_{4}\rangle\langle\psi_{2}\mid & \mid\psi_{3}\rangle\langle\psi_{3}\mid+\mid\psi_{4}\rangle\langle\psi_{4}\mid
\end{pmatrix}
\,,
\end{equation}
where now
\begin{eqnarray}
\label{SM27}
\mid\psi_{1}\rangle&=&\sqrt{\mathrm{C}}\cos\left(\lambda t\sqrt{\hat{n}+1}\right)\mid\alpha\rangle\;, \nonumber\\ \mid\psi_{2}\rangle&=&-i\,\sqrt{1-\mathrm{C}}\,\hat{V}\sin\left(\lambda t\sqrt{\hat{n}}\right)\mid\alpha\rangle\;, \nonumber\\
\mid\psi_{3}\rangle&=&-i\,\sqrt{\mathrm{C}}\,\hat{V}^{\dagger}\sin\left(\lambda t\sqrt{\hat{n}+1}\right)\mid\alpha\rangle\;, \nonumber\\ \mid\psi_{4}\rangle&=&\sqrt{1-\mathrm{C}}\cos\left(\lambda t\sqrt{\hat{n}}\right)\mid\alpha\rangle\,.
\end{eqnarray}
The atomic and field reduced density operators read
\begin{equation}
\label{SM28}
\hat{\rho}_{_{A}}=
\begin{pmatrix}
\langle\psi_{1}\mid\psi_{1}\rangle+\langle\psi_{2}\mid\psi_{2}\rangle & \langle\psi_{1}\mid\psi_{3}\rangle^{*}+\langle\psi_{2}\mid\psi_{4}\rangle^{*}\vspace{0.2cm} \\ 
\langle\psi_{1}\mid\psi_{3}\rangle+\langle\psi_{2}\mid\psi_{4}\rangle & \langle\psi_{3}\mid\psi_{3}\rangle+\langle\psi_{4}\mid\psi_{4}\rangle 
\end{pmatrix},
\end{equation}
and
\begin{equation}
\label{SM29}
\hat{\rho}_{_{F}}=\mid\psi_{1}\rangle\langle\psi_{1}\mid+\mid\psi_{2}\rangle\langle\psi_{2}\mid+\mid\psi_{3}\rangle\langle\psi_{3}\mid+\mid\psi_{4}\rangle\langle\psi_{4}\mid,
\end{equation}
respectively. In a similar fashion as the case previously described, we note that equation (\ref{SM29}) is nothing but the invariant defined in equation (\ref{SM3}), again for a four-level virtual atom. We can follow the procedure described above and plot in figure \ref{ASFPF05060} the field entropy, $S_{{F}}$, and field purity parameter 
$\xi_{_{F}}$ to show they again, as expected, have the same behaviour.

\begin{figure} 
\centering
\includegraphics[width=13cm]{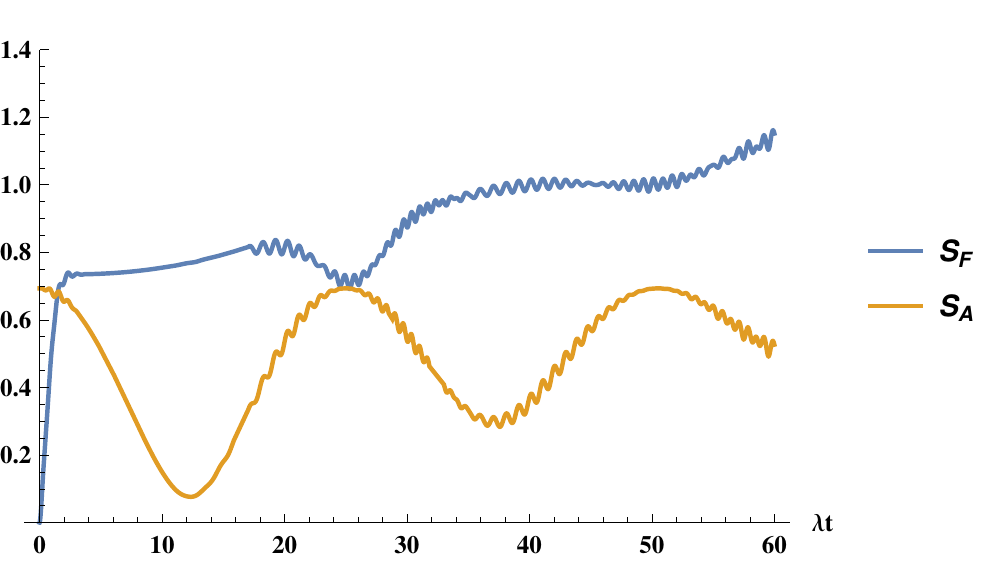} 
\caption{We plot the time evolution of the entropies, for the field, $S_{_{F}}$, and for the atom, $\S_{{A}}$ for the atom initially in an statistical mixture of excited nd ground states, with $\mathrm{C}=1/2$ and the field initially in a coherent state with $\alpha=4$.}
\label{ASFSA05060}
\end{figure}

Finally we show in figure \ref{ASFSA05060} the field and atomic entropies. In this case they show a completely different behaviour, unlike figure 2. 
\section*{Conclusions}
We have shown that in the atom field interaction, although the atom or  the field may be initially in mixed states, it is possible, by using the Araki-Lieb inequality and the concept of virtual (extended) atoms, to calculate the entropy of the field. Although the small system (in this case the two-level atom) continues to deliver the information of the big system (the field), its Hilbert space should be extended, in fact, doubled for us to be able to extract information about the complete system. 


\begin{thebibliography}{99}
\bibitem{Satya}  Satyanarayana, M.V.,  Rice, P.,  Vyas, R.   and  Carmichael, H.J. Ringing revivals in the interaction of a two-level atom with squeezed light. {\it  J. Opt. Soc. Am. B}  {\bf  6},  228--237 (1989).
\bibitem{Vidiella} Moya-Cessa, H.  and  Vidiella-Barranco, A. Interaction of squeezed states of light with two-level atoms. {\it  J. of Mod. Optics} {\bf  39},  2481--2499  (1992).
\bibitem{Phoenix3}   Phoenix, S.J.D. Wave-packet evolution in the damped oscillator. {\it Phys. Rev. A} {\bf  41},  5132--5138 (1990).
\bibitem{Abdel} Abdel-Kahlek, S., Berrada, K.  and Eleuch, H. 
Effect of the time-dependent coupling on a superconducting qubit-field system under decoherence: Entanglement and Wehrl entropy. {\it  Ann. of Phys.} {\bf  361},  247--258 (2015).
\bibitem{Neumann}  Von Neumann, J., Mathematical Foundations of Quantum Mechanics. Princeton University Press  (1955).
\bibitem{Banacloche} Gea-Banacloche, J. Atom- and field-state evolution in the Jaynes-Cummings model for large initial fields. {\it  Phys. Rev. A}  {\bf  44}, 5913--5931 (1991).
\bibitem{Vidiella2} Vidiella-Barranco, A.,  Moya-Cessa, H.  and Buzek, V. Interaction of superpositions of coherent states of light with two-level atoms. {\it J. of Mod. Optics} {\bf 39}, 1441--1459 (1992).
\bibitem{Phoenix2}  Phoenix,S.J.D.   and  Knight, P.L. Fluctuations and Entropy in Models of Quantum Optical Resonance. {\it  Annals of Physics} {\bf  186}, 381--407  (1988).
\bibitem{Barnett} Barnett, S.M., Beige, A., Ekert, A., Garraway, B.M. {\it et al.} Journeys from quantum optics to quantum technology. {\it Progress in Quantum Electronics} {\bf 54}, 19--45 (2017).
\bibitem{Araki}  Araki, H.   and  Lieb, E.H.  Entropy inequalities. {\it   Commun. Math. Phys.} {\bf  18}, 160--170 (1970).
\bibitem{23}{ Z\'u\~niga-Segundo, A., Ju\'arez-Amaro, R., Aguilar-Loreto, O.  and  Moya-Cessa, H.M.} Field's entropy in the atom-field interaction: Statistical mixture of coherent states. {\it Annals of Physics} {\bf  379}, 150--158 (2017).
\bibitem{10}{ Schmidt}, E. Zur Theorie der linearen und nicht- linearen Integralgleichungen I. {\it Math. Annalen} {\bf 63}, 433--476 (1907).
\bibitem{1}{Jaynes, E.T.  and  Cummings}, F.W. Comparison of Quantum and Semiclassical Radiation
Theories with Application to the Beam Maser. {\it Proc. IEEE.} {\bf 51},  89--109 (1963).
\bibitem{Gerry}  Gerry, C.C.  and  Knight, P.L. Introductory Quantum Optics. Cambridge University Press, 2005.


\end{thebibliography}
\end{document}